\documentclass[conference]{IEEEtran}
\usepackage{times}

\usepackage{graphicx}
\usepackage[numbers]{natbib}
\usepackage{multicol}
\usepackage[bookmarks=true]{hyperref}

\begin{document}

\title{Mitigating undesirable emergent behavior arising between driver and semi-automated vehicle}
\author{\authorblockN{Timo Melman\IEEEauthorrefmark{1}\IEEEauthorrefmark{2}\IEEEauthorrefmark{3}, Niek Beckers\IEEEauthorrefmark{1}\IEEEauthorrefmark{2} and David Abbink\IEEEauthorrefmark{2}}
\authorblockA{\IEEEauthorrefmark{2} Cognitive Robotics, Delft University of Technology, Delft, The Netherlands \\
\IEEEauthorrefmark{3} Chassis Systems Department, Group Renault, Guyancourt, France \\
T.Melman@tudelft.nl, N.W.M.Beckers@tudelft.nl, D.A.Abbink@tudelft.nl\\
website: www.delfthapticslab.nl \\
\IEEEauthorrefmark{1} Co-first author}}


\maketitle
\IEEEpeerreviewmaketitle

\begin{abstract}
Emergent behavior arising in a joint human-robot system cannot be fully predicted based on an understanding of the individual agents. Typically, robot behavior is governed by algorithms that optimize a reward function that should quantitatively capture the joint system's goal. Although reward functions can be updated to better match human needs, this is no guarantee that no misalignment with the complex and variable human needs will occur. Algorithms may learn undesirable behavior when interacting with the human and the intrinsically unpredictable human-inhabited world, thereby producing further misalignment with human users or bystanders. As a result, humans might behave differently than anticipated, causing robots to learn differently and undesirable behavior to emerge. 
With this short paper, we state that to design for Human-Robot Interaction that mitigates such undesirable emergent behavior, we need to complement advancements in human-robot interaction algorithms with human factors knowledge and expertise. More specifically, we advocate a three-pronged approach that we illustrate using a particularly challenging example of safety-critical human-robot interaction: a driver interacting with a semi-automated vehicle. Undesirable emergent behavior should be mitigated by a combination of 1) including driver behavioral mechanisms in the vehicle's algorithms and reward functions, 2) model-based approaches that account for interaction-induced driver behavioral adaptations and 3) driver-centered interaction design that promotes driver engagement with the semi-automated vehicle, and the transparent communication of each agent's actions that allows mutual support and adaptation. We provide examples from recent empirical work in our group, in the hope this proves to be fruitful for discussing emergent human-robot interaction. 
\end{abstract}

\begin{figure*}[!ht]
    \centering
    \includegraphics[width=0.9\textwidth]{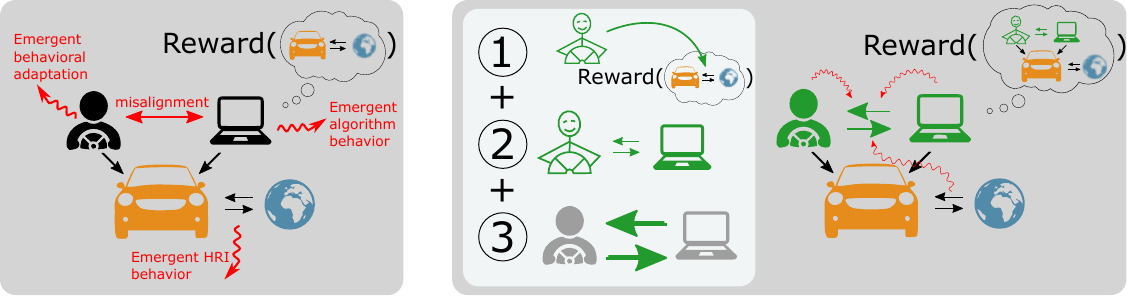}
    \caption{A schematic illustration of how undesirable emergent behavior may arise from the dynamic interaction between driver, self-learning algorithms, and the sAV in its environment (left panel). We propose a cybernetic interaction-based approach to mitigate this (right panel), combining three approaches: 1) leveraging human factors knowledge on driver's behavioral mechanisms and incorporate these in the sAV's algorithm, 2) including models of driver behavioral adaptation \textit{in interaction} with sAV's, and 3) resolving remaining misalignment and undesirable emergence through a driver-centered interaction design.}
    \label{fig:three-pronged-approach}
\end{figure*}

\section{Background on automotive use case}
Emergent behavior arises from the interaction between multiple agents in an environment, and are often unexpected and undesired (see Fig.~\ref{fig:three-pronged-approach}). This short paper focuses on avoiding undesirable emergent behavior arising from the interaction between two embodied agents in an automotive scenario: a \textit{driver} interacting with the \textit{algorithms} of a semi-automated vehicle (sAv) in a real-world environment. Predicting undesired emergent behavior is an extremely challenging task as it cannot result from understanding the vehicle in isolation of its driver, or vice versa. As such, the occurrence of emergent behavior depends not only on the driver but also on the automation algorithms, the interaction design, and the environment itself. 

Interestingly, vehicle automation algorithms are often based on data gathered in isolation, e.g. by mimicking naturalistic driving behavior. These algorithms may fail to generalize their learned behavior to scenarios that were not encountered before. The algorithms might even learn undesirable behavior and do not account for potential emergent behavior. This may cause misalignment that is annoying or even dangerous, hence the industry standard is to leave final responsibility with the driver. 
The driver may also behave in undesirable or even unexpected ways when interacting with the sAV. For example, drivers may not accept the automation due to misalignment or annoyance (disuse), or they may over-rely on the automation (misuse), which can lead to inappropriate responses (e.g., nodding off while driving in auto-pilot mode) or deskilling \cite{parasuraman1997_humans}. In human factors literature, disuse and misuse are referred to as \emph{behavioral adaptation}: undesirable adaptations in human behavior arising from the interaction between human and automation that mitigate the very benefits the automation aimed to realize. Disuse and misuse depend on the driver's attitude, the interaction design, the automation algorithms, and the dynamic environment. We, therefore, view behavioral adaptation as an important source of emergent behavior, arising from the interactive nature of driver and sAV.

\section{Mitigating undesirable emergent behavior}
This automotive example illustrates the point that to avoid undesirable emergent behavior in human-robot interactions, we need a three-pronged approach as illustrated in Fig.~\ref{fig:three-pronged-approach}. 
First, we should leverage human factors knowledge on driver's underlying behavioral mechanisms and incorporate these in the sAV's algorithm to mitigate misalignment at the source.
Second, human factors knowledge of driver adaptations \textit{due to the interaction with the sAV} needs to be included in the interaction design. This should avoid undesirable behavioral adaptations to the sAV. 
Third, all this should be grounded in an interaction design that limits the impact of misalignment \textit{when} it occurs, and allows for transparent communication of each agent's underlying intentions. 
In our group, we integrate these three approaches in an attempt to understand, observe, and mitigate undesirable emergent behavior. 

\subsection{Including behavioral mechanisms in the sAV algorithms}
For automated driving, the sAV's actions are generally selected by optimizing for a reward function that is defined in the design phase. To account for misalignment between driver preferences and sAV, algorithms can be used to infer the driver's (initially unknown) preferences, even while the sAV is already deployed (e.g., \cite{hadfield-menell2017_inverse,kuderer2015_learning}).
Literature provides several approaches to estimate and model naturalistic driving behavior, typically using \textit{rational} reward functions or inverse reinforcement learning techniques \cite{sadigh2016_information}. In our lab we adhere to the perspective of \textit{bounded rationality} in driving, which states that drivers tend to keep performance within \emph{acceptable, not optimal} states \cite{goodrich2000_satisficing}.
We have recently proposed a bounded-rationality risk-based reward function to propose a driver model from which risk-motivated adaptations in speed and trajectory naturally emerge \cite{kolekar2020_which}. 
This risk-based driver model was able to replicate a wide variety of such adaptations from literature, for example in response to changes in road width or oncoming traffic. It was also implemented in a test vehicle for a recent on-road study, providing human-like trajectory adaptations to a parked vehicle outside of the lane.

Still, no matter how well a risk-based reward function can capture naturalistic driving behavior, it will never do so perfectly for all drivers under all conditions. More importantly, a predefined reward function for the driver-sAV system does not necessarily distinguish between each agent's contributions to the driving task, whether driver behavior or algorithm behavior in isolation, or some form of cooperative driver-sAV behavior. As such, the sAV's algorithm does not account for undesirable emergent behavior such as when a driver - feeling supported - engages in more risk-seeking activities, resulting in an increase of the sAV's support, thereby progressively replacing the driver. In safety-critical scenarios outside the operational design domain of the sAV, such emergent behavior may prove detrimental. 

\subsection{Including predictions of driver behavioral adaptations}
A second approach is to include an empirical understanding of how driver behavior is shaped \emph{through the interaction} with the sAV (See 2 in Fig.~\ref{fig:three-pronged-approach}). 
Conceptual behavioral theories exist, called homeostasis theories, that aim to capture the underlying invariants (e.g., experienced task difficulty or risk) that govern driver adaptation to external changes. We recently replicated the three main theories to quantitatively compare their predictive capabilities into behavioral adaptation models \cite{melman2018_what}. We also used this knowledge to explain why drivers started speeding when supported by a haptic lane-keeping assistance system, thereby negating the safety and comfort benefits present at a fixed speed. Subsequently, we reduced haptic assistance above the speed limit, which subconsciously stimulated drivers to settle for a dynamic equilibrium around the speed limit, where haptic assistance provided safety and comfort \cite{melman2017_does}.
Such behavioral nudging may also occur \textit{between} different vehicles: e.g., sAV's can be designed such that they select their actions to elicit a desired behavioral adaptation from a human driver in another vehicle \cite{sadigh2016_planning}.

\subsection{Driver-Centered Interaction Design}
Because driver behavioral models and sAV reward functions will always be simplified representations of reality, approaches 1 and 2 will only reduce the occurrence and magnitude of misalignments and undesirable emergence, but not always prevent them. Therefore, a driver-centered interaction design is essential to cope with undesirable emergent behavior (from both sAV and driver) \textit{when they occur} (for a recent overview concerning the interaction design space based on shared control, see \cite{abbink2018_topology}). As long as drivers remain responsible, we advocate haptic shared control which makes misalignments tangible and correctable, thereby establishing mutual support between driver and sAV and the opportunity for mutual learning due to awareness of their dynamic capabilities and preferences.

\section{Symbiotic Driving for Beneficial Emergence}
When these three design approaches are combined, we speak of `symbiotic driving'. Recent test-track studies have demonstrated symbiotic driving outperforms sAV designs that have only improvements in human-like reward functions, or only improvements in interaction. Additionally, simulator studies show symbiotic driving avoids undesirable emergent behavior. In future work, we will explore symbiotic driving as a means to create the conditions for beneficial emergent behavior, also between multiple road users.

\section{Acknowledgment}
The work presented in this article was made possible by the Dutch Technology Foundation TTW (VIDI project 14127), which is part of the Netherlands Organization for Scientific Research (NWO).

\bibliographystyle{plainnat}
\bibliography{references}

\end{document}